\documentclass[10pt,a4paper,hidelinks,twocolumn]{article}

\usepackage[T1]{fontenc}
\usepackage[english]{babel}
\usepackage[utf8]{inputenc}

\usepackage{amsmath}

\usepackage{graphicx}
\usepackage[table,usenames,dvipsnames]{xcolor}
\usepackage{booktabs}
\usepackage{tabularx}

\usepackage{fancyhdr}
\usepackage{pdfpages}

\newtheorem{definition}{Definition}

\fancypagestyle{CISPA}{

\fancyhead[C]{\includegraphics[width=4cm]{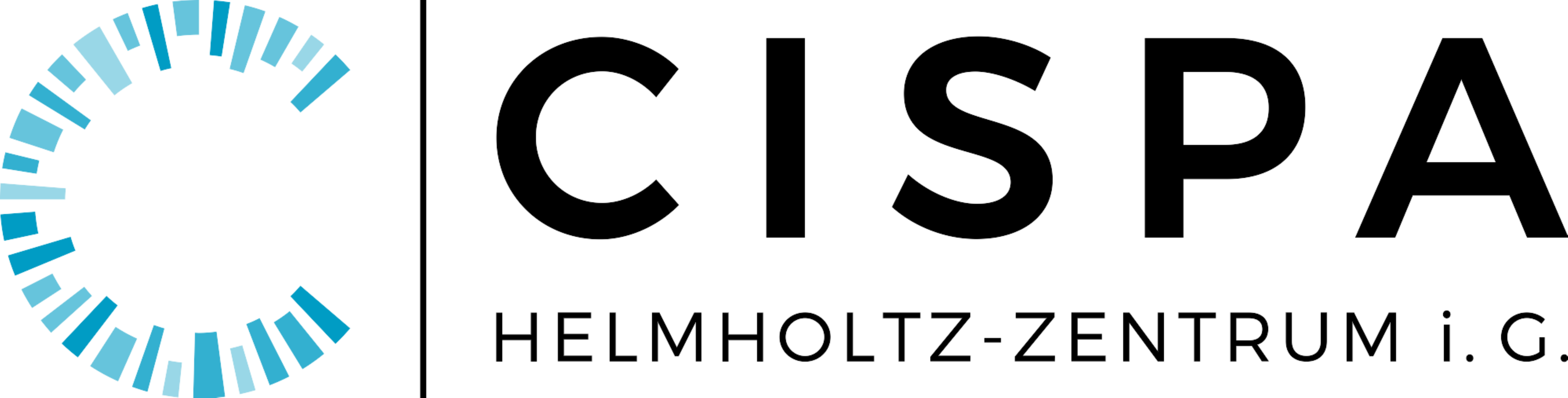}}
}
\usepackage{csquotes}
\usepackage{etoolbox}

\usepackage[nohead,includefoot,margin=2cm]{geometry}
\usepackage[compact,raggedright,small]{titlesec}
\usepackage[affil-it]{authblk}
\usepackage[runin]{abstract}
\usepackage{titling}
\usepackage{setspace}

\usepackage{multirow}
\usepackage{subcaption} 
\usepackage{caption}
\usepackage{tikz}
\usetikzlibrary{arrows}
\usetikzlibrary{positioning}

\usepackage{xspace}
\usepackage{hyperref}  
\usepackage[nameinlink,noabbrev]{cleveref}  
\usepackage[inline]{enumitem}
\usepackage{microtype} 
\usepackage{alltt}
\usepackage{float}
\usepackage{stfloats}

\usepackage[noend]{algpseudocode}
\usepackage{listings}
\usepackage{relsize} 

\definecolor{codegreen}{rgb}{0,0.6,0}
\definecolor{codegray}{rgb}{0.5,0.5,0.5}
\definecolor{codepurple}{rgb}{0.58,0,0.82}
\definecolor{backcolour}{rgb}{0.95,0.95,0.92}

\lstdefinestyle{mystyle}{
    commentstyle=\color{codegreen},
    keywordstyle=\color{magenta},
    numberstyle=\tiny\color{codegray},
    stringstyle=\color{codepurple},
    basicstyle=\footnotesize,
    breakatwhitespace=false,
    breaklines=true,
    captionpos=b,
    keepspaces=true,
    numbers=left,
    numbersep=5pt,
    showspaces=false,
    showstringspaces=false,
    showtabs=false,
    tabsize=2
}

\algblockdefx[switch]{Switch}{EndSwitch}%
[1]{\textbf{switch} #1}%
{}%
\algloopdefx{Case}[1]{\textbf{case} #1:}%
\algdef{SE}[DOWHILE]{Do}{doWhile}{\algorithmicdo}[1]{\algorithmicwhile\ #1}%

\def\|#1|{\textit{#1}}
\def\<#1>{\texttt{#1}}

\newcommand{\klee}{\textsc{Klee}\xspace}
\newcommand{\radamsa}{\textsc{Radamsa}\xspace}
\newcommand{\basilisk}{\textsc{Basilisk}\xspace}
\newcommand{\fuzzilisk}{\textsc{Fuzzilisk}\xspace}
\newcommand{\llvm}{\textsc{LLVM}\xspace}
\newcommand{\llvmir}{\textsc{LLVM~IR}\xspace}
\newcommand{\tgb}{Test Generation Bridge\xspace}

\newcommand{\sqlite}{SQLite\xspace}
\newcommand{\django}{Django\xspace}

\newcommand{\var}{\textit{var}}
\newcommand{\rvar}{\textit{rvar}}
\newcommand{\hrvar}{\textit{hrvar}}
\newcommand{\substr}{\textit{substr}}
\newcommand{\new}{\textit{new}}
\newcommand{\mse}{m^\textit{se}}
\newcommand{\rev}{\textit{rev}}

\usepackage{framed}
\newenvironment{result}{\begin{framed}\centering\it}{\end{framed}}
 
\title{Bridging the Gap between Unit Test Generation and~System~Test~Generation}
\date{\small (Dated \today)}
\author{Alexander Kampmann}
\author{Andreas Zeller}
\affil{\{alexander.kampmann, zeller\}@cispa.saarland \\
CISPA / Saarland University, Saarland Informatics Campus, Saarbr\"ucken, Germany}

\newcommand\BackgroundPic{
    \put(0,0){
    \parbox[b][\paperheight]{\paperwidth}{%
    \vfill
    \centering
    \includegraphics[width=\paperwidth,height=\paperheight]{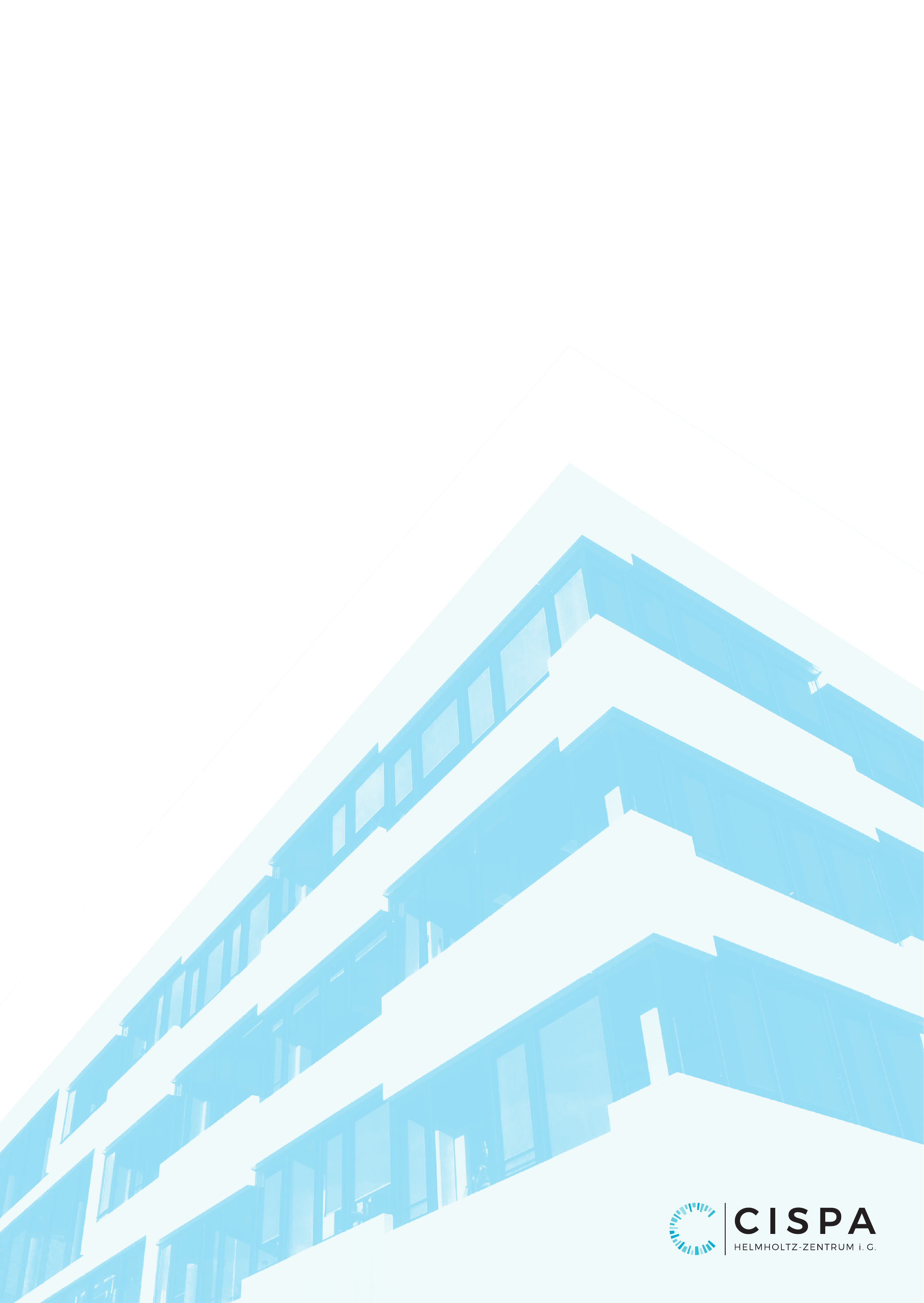}
    \vfill
}}}

\usepackage{utopia}
\usepackage{times}

\begin{document}
\AddToShipoutPicture*{\BackgroundPic}

\makeatletter
\renewcommand{\Authfont}{\normalsize\sffamily\bfseries}
\renewcommand{\Affilfont}{\normalsize\sffamily\mdseries}
\begin{titlepage}
\newcommand{\HRule}{\rule{\linewidth}{0.1mm}}
\centering
  \textsc{\LARGE {\fontfamily{Montserrat-TOsF}\selectfont CISPA Helmholtz-Zentrum i.G.}}\\[1.5cm]

  \vspace{2.4 cm}
  \HRule \\[0.2cm]
  {\huge\sffamily\bfseries \@title\par}
  \vspace{0.2cm}
  \HRule \\[1.5cm]

  {\sffamily \@author\par}
\vfill

\end{titlepage}

\makeatother
\setlength{\affilsep}{0.1em}
\addto{\Affilfont}{\small}
\renewcommand{\Authfont}{\normalsize}
\renewcommand{\Affilfont}{\normalsize}

\pretitle{\begin{center}\large\bfseries}
\posttitle{\end{center}}
\maketitle
\thispagestyle{CISPA}

\begin{abstract}
Common test generators fall into two categories.  Generating test inputs at the \emph{unit} level is fast, but can lead to false alarms when a function is called with inputs that would not occur in a system context.  If a generated input at the \emph{system} level causes a failure, this is a true alarm, as the input could also have come from the user or a third party; but system testing is much slower.

In this paper, we introduce the concept of a \emph{test generation bridge}, which joins the accuracy of system testing with the speed of unit testing.  A \tgb allows to combine an arbitrary system test generator with an arbitrary unit test generator.  It does so by \emph{carving parameterized unit tests} from system (test) executions.  These unit tests run in a context recorded from the system test, but individual parameters are left free for the unit test generator to systematically explore.  This allows symbolic test generators such as KLEE to operate on individual functions in the recorded system context.  If the test generator detects a failure, we lift the failure-inducing parameter back to the system input; if the failure can be reproduced at the system level, it is reported as a true alarm.  Our BASILISK prototype can extract and test units out of complex systems such as a Web/Python/SQLite/C stack; in its evaluation, it achieves a higher coverage than a state-of-the-art system test generator.
\end{abstract}

\section{Introduction}
\label{sec:introduction}



\begin{figure}
	\includegraphics*[width=\linewidth]{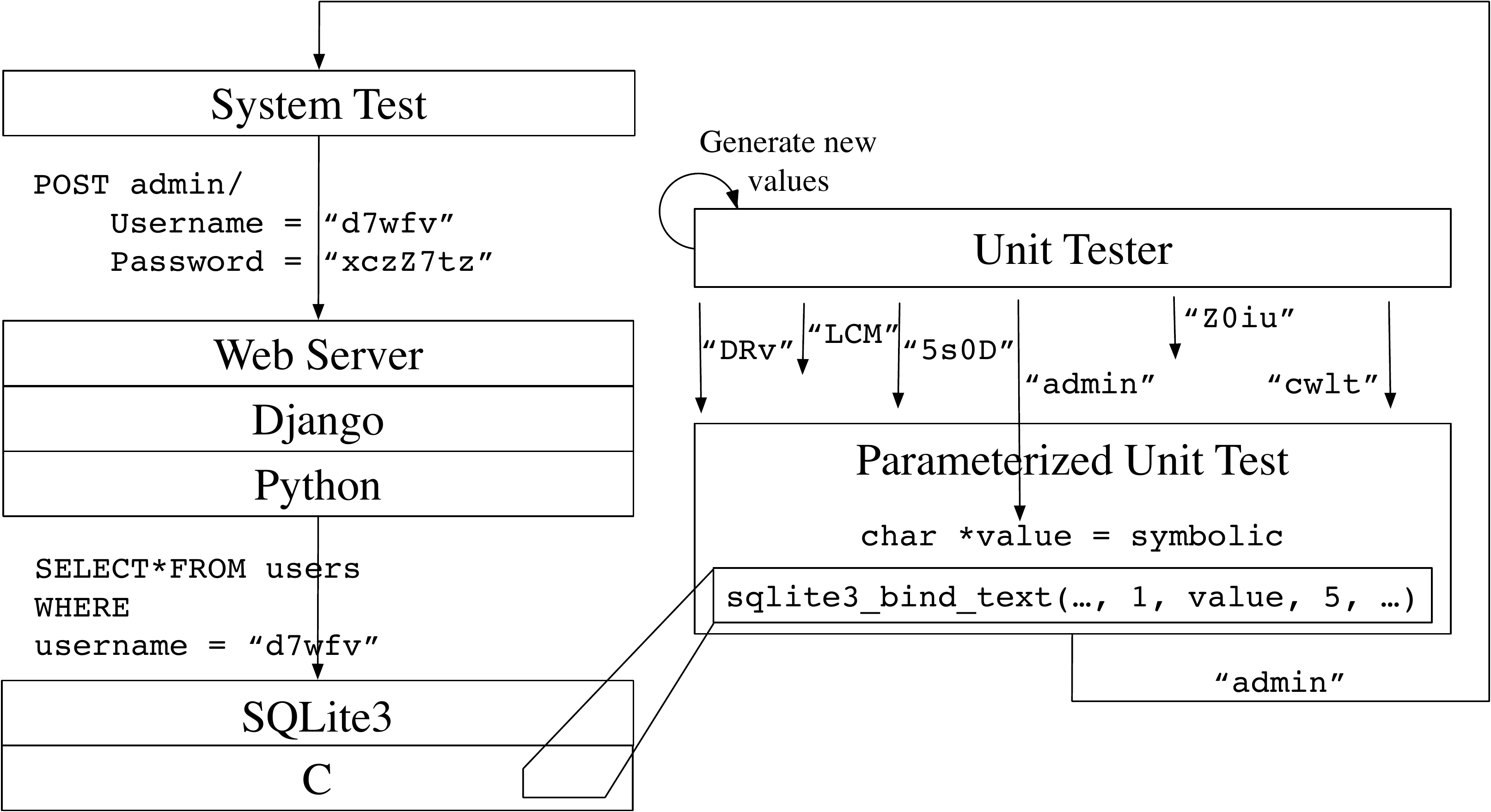}
	\caption[Overview of our approach.]{Applying our approach to a multi-layered example.  
		Starting from a system test, a HTTP login request, data is processed in the Django framework (Python), passed to the pysqlite python extension, and further on to \sqlite (both written in C).  We generate a parameterized unit test from the \sqlite layer, allowing a unit test generator to produce arbitrary many input values.  Input values that lead to failures are lifted back to system tests.}
	\label{fig:example}
\end{figure}

Software test generation can take place at two levels.  Generating tests at the \emph{unit} level calls functions with synthesized arguments.  Generating tests at the \emph{system} level feeds synthesized inputs into a system.  Both approaches have their advantages and disadvantages, as we can easily illustrate.  Consider the system shown in \Cref{fig:example}: A web server invokes \django Python code embedding a SQL database, which in turn contains C functions.  A web test generator can analyze the (HTML) user interface and interact with the system through a remote-controlled browser.  The generated inputs result in Python function invocations, which in turn produce SQL commands; these are interpreted by the embedded SQL library, which at the lowest layer consists of C functions interacting with the operating system.  

This stack comes with huge execution times, which restricts how many system tests can be executed in a given time frame. System testing is slow.  The alternative to system testing is unit testing.  Since unit tests invoke individual functions directly, they are faster than system tests by construction.  \emph{Generating} unit tests, however, can be hard, as one has to synthesize function arguments that actually satisfy the assumptions of the function called.  As an example, consider one function of \sqlite, say \<sqlite\_bind\_text()>.  This function, which 
binds a value to a prepared SQL statement, takes five parameters---its full signature is
\begin{lstlisting}
int sqlite3_bind_text(sqlite3_stmt *stmt, int id, const char *value, int length, void (*destruct)(void *));
\end{lstlisting}
The awkward \<destruct> argument is a pointer to a function used to dispose of the string after \sqlite has finished with it. 
A test generator has to synthesize correct values for all five arguments.   The greatest challenge is to produce a \<stmt> argument---the \<sqlite3\_stmt> structure represents the internal structure of a parsed SQL statement.  In the absence of a formal specification, there is no way for a test generator to determine what makes a valid \<sqlite3\_stmt> structure.  A test generator can \emph{create} such a structure by invoking a \<sqlite\_prepare\_v2()> parsing function---but this function requires, among others, a valid SQL statement.  How would a unit-level test generator ever synthesize these?

Practical unit-level test generators (e.g. \cite{Fraser2011evolution}, \cite{pacheco2007randoop})  address the problem with a ``best effort'' approach---synthesizing as much data as possible and feeding this into the function under test.  If the function fails, however, one cannot know whether this is because of the generator failing to provide correct and meaningful inputs (after having synthesized an incomplete \<sqlite3\_stmt>, for instance), or because of the function actually having a bug\cite{gross2012search}.  This problem of \emph{false alarms} does not occur at the system level, because the system must be prepared for invalid inputs. The end user could also provide those.   But then again, system testing is slow.

\begin{figure}
	\includegraphics*[width=\linewidth]{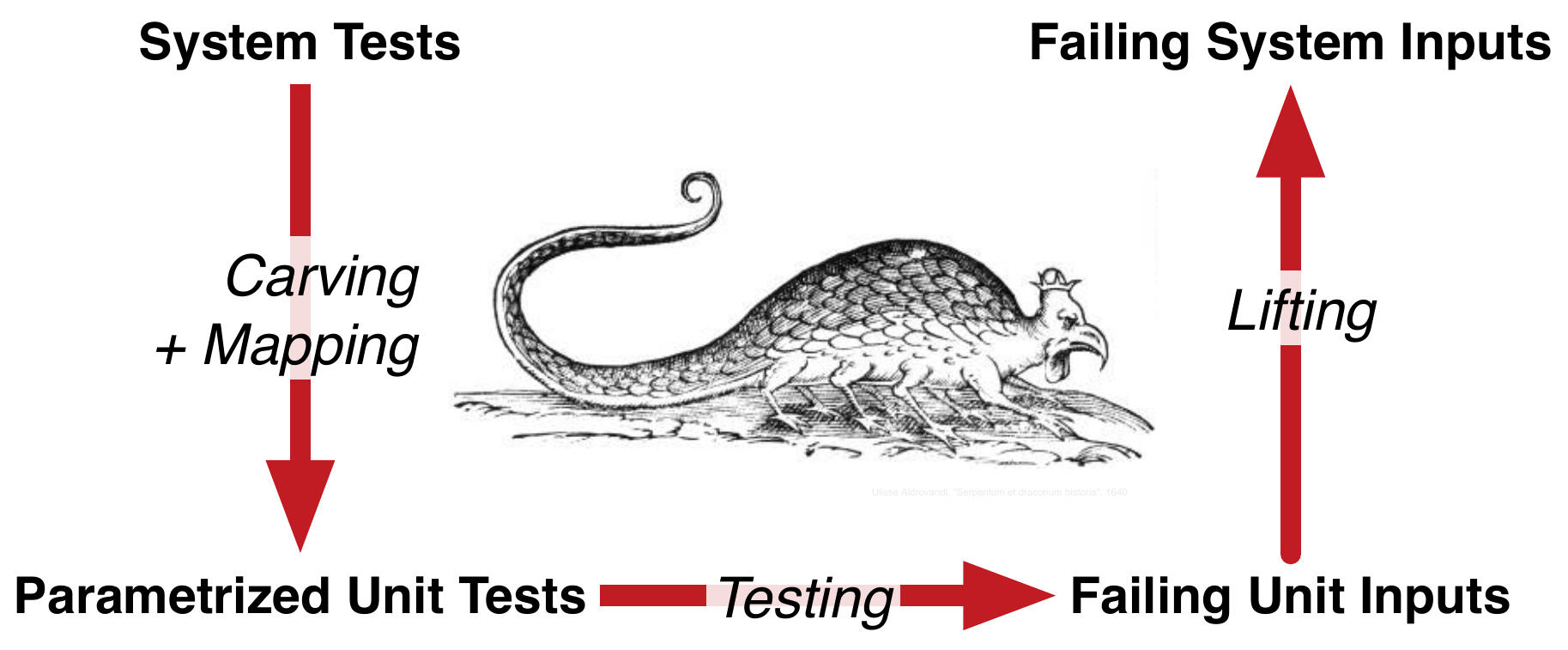}
	\caption[Overview of our approach.]{The \tgb works by \emph{carving} unit tests from a system test execution, \emph{testing} on the unit-level and \emph{lifting} discovered failures back up to the system level.}
	\label{fig:overview}
\end{figure}

In this paper, we introduce an approach which joins the accuracy of system testing with the speed of unit testing.  A \emph{test generation bridge} is a two-way connector joining existing \emph{system-level} test generators with an existing \emph{unit-level} test generator.  As shown in \cref{fig:overview}, a \tgb propagates information from system level to the unit level and back again in these three steps:

\begin{description}
	\item[Carving and Mapping.] 
	During system test generation, the \tgb carves out individual functions as unit tests.  Each such unit test comes with global variables set to the values recorded during execution. In carving, we identify variable values (including function arguments) that also occur (as strings or integers) in the system input and create a mapping between those variables and the input fragments.  Mapped function parameters are replaced with symbolic values, allowing for parametrized unit testing. In our example, we carve the invocation of \<sqlite3\_bind\_text()>, and make the third parameter symbolic, because it is identical to a value used in the system test.
	
	\item[Testing.]
	On the extracted parametrized unit tests, a \emph{unit test generator} can now determine function arguments that cause the function to fail. As the units are small, the test generator can make use of expensive techniques such as symbolic execution; as function execution is fast, the test generator can execute the function several times.  In our example, a single execution of \<sqlite3\_bind\_text()> is 45$\times$ faster than a single system test.
	
	\item[Lifting.]
	Even in a recorded context, unit testing with synthesized inputs can result in false positives.  To check whether inputs causing failures at the unit level  are also feasible at the system level, the \tgb \emph{lifts} them back into system inputs, using the previously established mapping: If such a system test fails, we have found a true failure. In our example, we generate a new system test, using the value that unit level analysis found. 
\end{description}

By carving units, testing them in context, and lifting solutions back to the system level, a \tgb thus \emph{integrates} existing system and unit test generators, bringing together their respective strengths, features, and maturity.  The bridged gap between system and unit test generator can be arbitrarily large and cross multiple language and abstraction barriers, including the Web/Python/SQLite/C stack above.


The remainder of this paper is organized as follows.  \Cref{sec:example} demonstrates the carving, mapping and lifting steps on our Web/Python/SQLite/C example, and afterwards \Cref{sec:bridge} shows their details and interplay.
In \Cref{sec:implementation}, we present our proof-of-concept implementation, which carves unit tests out of \<C> programs, in two configurations: \basilisk combines \radamsa as a system-level fuzzer with \klee as a unit-level analysis. For \fuzzilisk, we combine \radamsa on the system-level with our own unit-level fuzzer. 
In \Cref{sec:evaluation}, we evaluate \basilisk on a series of C~programs.  We find that carving parameterized unit tests can yield more coverage in less time when compared against system-level testing; these savings are multiplied when focusing the testing effort on a subset of the program.  \Cref{sec:conclusion} closes with conclusion and future work.

\section{Example}
\label{sec:example}

In our running example, we consider HTTP requests against a web application backend, which is realized with \django\cite{django}. The \django framework provides the ability to define a model, which is automatically mapped to a database, controllers, which operate on the model as a result of user interactions (HTTP requests) and templates, which render the result of a controller to a HTML page. \django also auto-creates an administration interface, which allows to create, retrieve, update and delete (CRUD) model objects. \django uses an embedded \sqlite database for storage. 

For our example, we created HTTP login requests for the automatically generated administration interface of a \django web app. Our tool had the ability to choose the user name and password to try. Such a HTTP request goes to the \django web server, written in python, and is processed in the login controller. This controller generates SQL queries, which translate to look ups in the underlying \sqlite database. The database is written in C. 

\subsubsection{Carving and Mapping}

We first applied \radamsa, a system test generator. \radamsa chose random values for user name and password, as an example, the tool may try \|d7wfv| as a user name and \|xczZ7tz| as a password. 

We carved unit tests from the \sqlite database, restricting ourselves to the \<sqlite3\_bind*> family of functions. Executing a query in \sqlite is a three step process. First, a query is \emph{prepared}, with the \<sqlite3\_prepare\_v2()> function. Preparing a query pre-compiles the SQL statement into object code, leaving placeholders for all values used in the query. Second, \<sqlite3\_bind*> functions are used to replace those placeholders with values. There are 16 \<sqlite3\_bind*> functions in total, one example is \<sqlite3\_bind\_text()>, which we have shown in \Cref{sec:introduction}. It has five parameters, only one of which, the actual value, is dependent on user input. 

Lastly, \<sqlite3\_step()> actually executes the query, returning the constants \<SQLITE\_ROW> if there is another value in the database or \<SQLITE\_DONE> if all values matching the search parameters where found.  

In our example, we observe a call to \<sqlite3\_bind\_text()> for the string \|d7wfv|. Also, we record the value of all global variables, and all arguments to \<sqlite3\_bind\_text()>. We observe that the string parameter of \<sqlite3\_bind\_text()> is \|d7wfv|, identical to the user name we gave.

Now, we generate a unit test which re-creates the state the program was in when \<sqlite3\_bind\_text()> was called. This recreates the prepared statement object, which, in the original execution, was initialized by a call to \<sqlite3\_prepare\_v2()>. 

\subsubsection{Testing}

Our mapping records that the text parameter to \<sqlite3\_bind\_text()> is identical to a system-level input, the user name. Thereby we make this value a parameter of the unit test. Now, we can run a unit level test generator. In our experiments, we used \klee\cite{Cadar2008klee}, as well as a fuzzer we wrote ourselves. Both tools may not mutate the context, but only the values we marked as parameters. In this case, this is the user name. 

\klee applies symbolic execution. It executes the program under test and records all operations performed on individual values, and generates constraints from these. Then, it uses an SMT solver to generate values which \emph{solve} those constraints. The solutions are used as new test inputs. 

We could not apply this on the system level, because \klee would not be able to handle HTTP requests or the python part of the program under test. It can, however, handle the C part. When it analyses the carved unit test, it suggests the values \|admin| and \|q|. For both strings, it reports that the unit test executes some previously uncovered code if they are used instead of \|d7wfv|. 

\subsubsection{Lifting}

We know, from the mapping stage, that the string \|d7wfv| re-appeared on unit level, and was replaced with \|admin| and \|q| by the unit analysis tool. So we just replace \|d7wfv| with \|admin| and \|q| in the system test. This yields two new system tests which performs the same HTTP requests, but this time with \|admin| or \|q| respectively as a user name. 

When using \|q| as a user name, the system test shows the same behavior as before. 
Reporting \|q| as a new input is a \emph{false positive}. The \tgb recognizes this, and does not report a system test containing \|q| to the software developer. 

However, \|admin| is in fact the correct user name. The system test with this value achieves additional code coverage, namely the code path which, after discovering a correct user name, checks the password. \klee found the value in the carved context, which contains the entire database, and suggested it as a better test input at unit level. This shows that we can in fact discover good input values at the unit level and lift them to the system level. 

\section{The Test Generation Bridge}
\label{sec:bridge}

The \tgb consists of the following steps:

\subsubsection{Carving and Mapping}

Carving extracts unit-level interactions from a system test and generates a unit test, based on those interactions. The carving tool also has to provide a \emph{partially reversible mapping} of unit-level values to system-test inputs.

\subsubsection{Testing}

The unit tests are turned into parameterized unit tests by replacing all \emph{reversible} values with a parameter. Testing then determines values for those parameters which make the test fail. 

\subsubsection{Lifting}

Lifting takes the values that were generated by testing and \emph{lifts} them back up to the system level, thereby checking whether values which cause a failure at the unit level do so at the system level as well. 

In our formalization of the \tgb, we will often use the phrase ``an execution discovers a test goal $g$''. The exact meaning of this depends on the testing objective. If the objective is branch coverage, a test goal $g$ is a branch, and it is discovered by executing (covering) it. If the objective is bug finding, a test goal $g$ is discovered if a program execution provides evidence for the existence of a bug. 





\subsection{Carving}

Elbaum et al.\cite{elbaum2009carving} presented carving as a way to generate unit tests out of system tests.  Elbaum's carving technique is non-parameterized; that is, the carved tests reproduce exactly the same calls seen during system testing.  In contrast, our technique carves \emph{parameterized} unit tests, allowing a unit-level testing tool to provide new values for function parameters.

Conceptually, a system test can be represented as the execution path through the program that is taken when it processes the test inputs. Carving slices this execution path horizontally. It picks a subpath of the execution path and uses this subpath as a unit test. 
\begin{definition}
	We call the selected subpath the \emph{test path}. The \emph{execution prefix} is the prefix of the execution path before the test path, and the \emph{execution suffix} is the part of the execution path after the test path. 
\end{definition}
During the execution of a test path $t$, variable values will be used to inform branching decisions. 
\begin{definition} 	
	Let~$C_t$ map all local variables in the function that contains the start point as well as all global variables to their values. If $v$ is a local or global variable, we use~$C_t[v]$ to denote the value~of~$v$ at the start of~$t$. We call $C_t$~the \emph{context} of~$t$. 
	We call $\var(C_t)$ the set of all variables which exist at the start point of a test~path~$t$.  
\end{definition}

This definition of a context does not cover the entire state of the program. It has no information about the call stack, that is, local variables in functions deeper in the stack as well as the structure and contents of the stack itself. Also, if the program under test has files opened or uses other external resources, those will not be fully reflected in the context. 

Assuming that the program does not use anything but variable values to inform branching decisions, the test path is fully defined by the start point and the context at the start point. Therefore, such a test path can be re-executed if the program is brought into the same context. That is, all variables are populated as recorded in the context and the program execution is started (and later, stopped) at the correct instruction. 

We consider this assumption reasonable, because programs rarely depend on the structure or content of the call stack. 
And while the context does not cover external resources, values that were read from an external resource and are stored in a variable at the start point are part of the context. 

\begin{definition}
	We call two contexts~$C_1$ and~$C_2$ \emph{compatible}, if~$\var(C_1) = \var(C_2)$. 
\end{definition}

Obviously all test paths can be executed in their original context~$C_t$. Executing a test path in a compatible context~$C_2$ means invoking the same start point with different variable values. This may not be possible if~$C_2$ is not compatible with~$C_t$, variables which are used within~$t$ might be missing. If the start point of a test path $t$ is executed in a context different from $C_t$, branching decisions within the test path may have a different outcome, forcing the execution along a path different from~$t$. 

\subsection{Mapping}

\begin{definition}
	In addition to the context, we define the \emph{set of system-level inputs} $S$. This set contains all strings\footnote{Conceptually, those may be numbers, images or more complex structures, but they are usually passed into the system as byte strings. } which the program under test reads from external resources or receives as command line parameters. 
\end{definition}

Let~$S$ be a set of system-level inputs, $t$~a test path which occurs in an execution of the program under test with the inputs~$S$, let~$C_t$ be the context for~$t$. We require the carving tool to provide a mapping~$m_{S,C_t}$ from context values to system-level inputs. The mapping needs to be \emph{partially reversible}. 

For sake of readability, we will sometimes write $m$~instead~of~$m_{S,C_t}$. Despite the shortened notation, a mapping is always accompanied by a set of system-level inputs, a context and a test path. 

\begin{definition}
	We call a mapping $m_{S,C_t}$ \emph{reversible for a context $C_2$}, if $C_2$ is compatible with $C_t$ and executing the system-level inputs $S_2 = m_{S, C_t}(C_2)$ leads to an invocation of the start point of~$t$ in context~$C_2$.
	
	Let $C_\rev(m)$ be the set of all contexts $C$ such that $m$ is reversible for $C$. We call a mapping \emph{partially reversible}, if $C_\rev(m)$ is not empty (that is, if it is reversible for some contexts). 
\end{definition}

The \tgb requires such a partially reversible mapping, and, in practice, works best if $C_\rev(m)$ is large. 

\begin{definition}
	We call a variable $v$ \emph{reversible} with respect to $m$ if there is a context $C_v$ such that $C_v[v] \neq C_t[v]$ and $C_v \in C_\rev(m)$. We call the set of all reversible variables $\rvar(m)$. 
\end{definition}

Note that we only require the existence of one such context, there can be any number of contexts $C_v'$ such that $m$ is not reversible with respect to $C_v'$. 

Obviously, it is hard to provide such a partially reversible mapping. A branching decision based on~$s \in S$ somewhere in the execution prefix of~$t$ may mean that the start point of~$t$ is never reached in an execution of the program under test with the inputs~$S_2$. This could be solved with symbolic execution: If the path constraints for an execution are known, a solver might be able to provide the values for~$m_{S, C_t}(C_2)$, essentially reversing the constraints. 

However, those constraints might be difficult to solve. In our running example, the password is part of the system inputs. The password is hashed and salted before it is send to the database. Now, even if we can identify $v$ such that~$C[v]$ is the hashed password, we would be required to reverse the hash function. Cryptographic hashes are designed to make this computationally impossible. 

To mitigate this issue, implementations are only required to provide a set $\hrvar(m)$ such that the overlap $\hrvar(m) \cap \rvar(m)$ is large. That is, they should provide the most reasonable "best guess" at what $\rvar(m)$ looks like. 

Within the running example, $S$ contains \verb|d7wfv| and \verb|xczZ7tz|, the randomly chosen values for user name and password. $C$ contains the values for all parameters to \<sqlite3\_bind\_text>, those are \<value = "d7wfv\textbackslash0", length = 6, id = 1, destruct = NULL>, and a complex heap object for \<stmt>. The tool recognizes that $C_t[\texttt{value}] \in S$, and thereby gives $\texttt{value} \in hrvar(m)$. 

\subsection{Testing}
\label{sec:texting}

We build a unit test from~$t$~and~$C_t$ such that the unit test invokes the start point of~$t$ in the context~$C_t$. Then, we mark the variables $v \in \hrvar(m)$ as \emph{parameters}, allowing a unit-level testing tool to provide new values for those parameters. 

While the original unit test necessarily follows the execution path~$t$, the parameterized test does not need to do so. If there are branching decisions directly or indirectly based on the values of some~$v \in \hrvar(m)$, the parameterized test may take a different path, depending on the values that the unit-test input generator chose for those variables. The unit-level test generator will try to generate inputs which reach some test goal $g$ that was not discovered by the original test. 

Still, the parameterized unit test can only reach parts of the program which are reachable from the start point of~$t$. Thereby the amount of program code which the unit-test input generator needs to analyze is much smaller than the entire program under test. This enables us to use techniques which do not scale to large programs on the unit-level end of the bridge. 

We then generate a new context~$C_{\new}$. For this context, we choose $\var(C_{\new}) = \var(C_t)$. For all $v \in \hrvar(m)$, the parameters of the unit test, we use the values that were provided by the unit-level analysis. For all $v' \not\in \hrvar(m)$, we choose $C_{\new}[v'] = C_t[v']$. 

\subsection{Lifting}
\label{sec:lifting}

Let's assume that~$C_{\new}$ is reversible by~$m$. This means invoking the program under test with the system-level inputs~$S_{\new} = m_{S, C_t}(C_{\new})$ leads to an invocation of the start point of~$t$ in context~$C_{\new}$. But then, this execution will also discover~$g$. So~$S_{\new}$ is a set of test inputs which discover test goal~$g$. We call the process of applying $m$ to generate such a system-level test input $S_{\new}$ \emph{lifting}. 

As mentioned before, several things can go wrong:
\begin{enumerate}
	\item The mapping is \emph{partially reversible}. If $C_{\new} \not\in C_\rev(m)$, an execution of the program under test with the inputs~$S_{\new}$ may not trigger an execution of the start point of~$t$ in~$C_{\new}$. 
	\item The \emph{context} may have been insufficient. The test goal~$g$ might represent a behavior that is triggered because the context was insufficient. For example, a file might not have been opened for the parameterized unit test. This triggers error handling code, and thereby a test goal. However, an execution of the program under test with the inputs~$S_{\new}$ would most likely not trigger this, as the file would be opened by the system. 
	\item The \tgb assumes that the program under test is deterministic and that program behavior depends on nothing but the inputs in~$S$. If this assumption does not hold, the behavior of $S_{\new}$ is unpredictable.
\end{enumerate}

All three issues lead to the same outcome. If we execute the program under test with the inputs~$S_{\new}$, the test goal~$g$, which was discovered by the unit test, is not discovered in the system test. The prediction of fulfilling~$g$ with~$S_{new}$ is a \emph{false positive}. The lifting stage of the \tgb performs this execution and thereby can avoid to bother human software developers with false positives. Due to this validation within the lifting step, we can afford imprecision in any other step of the \tgb. Incorrect results will just be filtered later. 

It might happen that the execution of the program under test with the system-level inputs~$S_{\new}$ triggers some other test goal. If there is a branching decision in the execution prefix of~$t$ which is based on some~$s \in S$, and this value is different in~$S_{\new}$, the branching decision may have a different outcome, and~$t$ will never be reached. However, if the test goal was branch coverage, this branch may have been a test goal, leading to a situation where executing the program under test with the inputs~$S_{new}$ covers a test goal~$g'$ different from the test goal~$g$ that would be covered by executing~$t$ in~$C_{\new}$. We observed this several times for branch coverage as a testing goal, but it is less likely to happen if the testing objective was bugs. 

\section{Implementation}
\label{sec:implementation}

We built two versions of the \tgb:
\basilisk combines \radamsa as a system tester with \klee on the unit level.
\fuzzilisk combines \radamsa as a system tester with a self-written, unit-level fuzzer. 
Within both implementations, we define the test path by isolating a method call. That is, the method invocation is the start point and the method return is the end point of a test path. Carving then generates a unit test which initializes all variables to the observed values and calls the selected method with the observed parameter values. 

\subsection{Carving}

\label{sec:carving-implementation}


Our \basilisk prototype implements carving based on \llvm\cite{lattner2004LLVM}. \llvm~provides a compiler, \emph{clang,} which compiles C code to an intermediate representation (\llvmir). Clang also compiles \llvmir~to machine-code. 

The \llvmir intermediate representation is designed to be used by compiler optimizations, that is, static analysis of the code. It removes all the syntactic sugar that was added to C for the sake of human developers, so it is hard to read for humans, but better suited for automated analysis. At the same time, it preserves type information, which makes an analysis simpler than analyzing machine code directly. 

Our carving approach works in two phases:
\begin{enumerate} 
	\item It statically instruments the \llvmir~code, inserting probes which report all method invocations including the parameters, as well as all writes to global variables. 
	\item During execution, the probes write the observed values at those points to a trace file.
\end{enumerate}
For primitive types such as numbers, observed values can be written directly.  Other data types pose more challenging situations, as listed below. 

\subsubsection{Pointers}

In \llvmir, as in C, a pointer is no more than a memory address. There is no information about the length of the memory segment that a pointer points to. But then, what should we dump out? 

If we dump out only the byte that the pointer points to, we do not see the contents of the remainder of the memory area. Keep in mind that pointers are often used as arrays, e.g. a \<i8*> pointer, \llvmir~for a \<char *> pointer, is often used to point to a string, an array of characters in memory. In this case, dumping just one byte would give only the first character of the string. 

At the same time, we cannot just dump arbitrary amounts of data. We would dump data that does not belong to this variable, and in some cases, accessing memory past the end of an allocated memory segment could even trigger a segmentation fault, and thereby crash the program under test. 

To solve this problem, we maintain a map that records, for each pointer, the length of the memory segment it points to. Keep in mind that pointers can be calculated from other pointers. If a pointer \<a> points to a memory segment of length 10, \<b = a + 5> gives a pointer \<b> which points to a memory segment of length 5: the second half of the segment pointed to by \<a>. The map needs to be able to identify \<b> as pointing into the memory segment at \<a>, and calculating the remaining length. This allows us to find out how much memory should be dumped for a pointer. 

We monitor all memory allocations to populate the map. This includes calls to \<malloc> (heap memory), the \llvmir \<alloca> (stack memory) instruction, constants (allocated in the \<.text> section of the binary) and arguments to \<main> (technically stack memory, but placed by the operating system instead of \llvmir's \<alloca>). For the \django example, we also monitor the \<PyString\_As*> functions, which are used to move data that was allocated within the python interpreter to the C part of the subject. 



\subsubsection{Strings}

For strings, one might assume that we can rely on the fact that they are zero-terminated, that is, the last character will be zero. Unfortunately we can not. Zero-termination is a convention, not a rule. Programmers may also decide to accompany their \<char*> variables with an integer variable, holding the length of the string. Also, we encountered several cases where a bitset was stored in a \<char *>. In a bitset, there may be relevant data behind a zero byte. So assuming zero-termination in those cases means that we would loose data. We therefore decided to handle \<char *> in the same way as any other pointer type. 


\subsubsection{Structs and Unions}

Struct types and union types are derived types. For a struct, that means that the struct consists of several values of different types, written to memory one after another. 
A union describes several options on how to interpret the same memory segment, e.g. 4-bytes of data may either be a single int32 value, or a four-letter string. 

We dump structs by handling each field recursively. Unions are compiled to structs and bitcasts (reinterpreting bytes in memory as some other type) in \llvm, so we don't need to think about how to handle them, \llvm already did.

\subsubsection{Extensive lengths}

Large structure, for example large arrays or large connected heap structures, take a long time to dump. Also reconstructing them will be time consuming, which leads to slow unit tests. As our evaluation will show, slow unit tests are a problem for our approach. 

We solved the problem with a size limit. If a structure or individual array is too large, dumping will be aborted, and the program under test just continues. This means that the context may be incomplete. However, as the lifting step filters false alarms, this is not a major threat to our approach.



\subsubsection{External Resources}
If the program under test had a file open when $f$ was invoked, reconstructing variable values and heap structures will not be sufficient. Calls to \<read()> or \<write()> on the file will fail.  Similar situations may occur with other resources, such as locked mutexes or open network connections.  

A perfect solution would have to deeply interact with the operating system (and the system environment) to perfectly preserve and reconstruct states, which is out of our scope. We therefore just ignored the issue. It may lead to false alarms in the unit tests, but our lifting step will filter those. 

\subsubsection{Writing to Global Variables}

We dump values for global variables as they are written, that is, every time a \<store> instruction is executed. However, we only want to dump data when \<store> in fact changes global memory. Therefore, when our instrumentation encounters a \<store> instruction, it traces back the pointer operand. If the pointer is a global variable or derived from a global variable, the instrumentation adds code to dump the global variable after the \<store>. 


%



\subsection{Mapping}

In this section, we describe the mapping~$\mse$, which is used in our implementation. $\mse$ is based on the idea of using string comparisons. This means that $\mse$ can be calculated quickly, but is more imprecise than, e.g. a mapping based on tainting. 

For each carved unit test, we have the set of system-level inputs, $s \in S$, and the test context $C_t$. 

Let~$\substr(s) \subseteq \var(C_t)$ be all variables~$v$ such that~$C_t[v]$ is a substring of~$s$. If~$C_t[v]$ is an integer, we say~$v \in \substr(s)$ if the decimal encoding of~$C_t[v]$ is a substring of~$s$. 

Let $S_u = \left\{ s \vert s \in S \wedge \forall_{v \in C_t} s \not\in substr(v) \right\}$ be the set of all system-level inputs which are not related to a variable in $C_t$. Those will not be affected by the mapping. 

Let $S_c = \left\{ c_{\new}[v] \vert v \in \var(C_t) \wedge \exists_{s \in S} v \in substr(s) \right\}$ be the set of the new values for variables $v$, for which the mapping found a substring match in $S$. 

When we apply the mapping, it will be $S_{\new} = S_u \cup S_c$. The mapping does not change values that do not have a substring match ($S_u$), and changes values where a substring match could be found~($S_c$). 

For using $\mse$ in the \tgb, a value for $\hrvar(\mse)$ is required. Those variables will be parameters to be discovered by the unit-test analysis. We define $\hrvar(\mse)$ as
$$
\hrvar(\mse) = \left\{ v \vert v \in \var(C_t) \wedge \exists_{s \in S} v \in substr(s) \right\}
$$ 

It remains to see whether $\mse$ is partially reversible.  Assume that we have a context $C_{\new}$ such that $C_{\new}[v] \neq C_t[v]$. We calculate $S_{\new} = \mse(S, C_{\new})$. Let $C_{\text{obs}}$ be the context at the start point $t$ that we observe in the execution of $S_{\new}$. 
This results in four possible outcomes:
\begin{enumerate}
	\item $C_{\text{obs}}$ does not exist because $t$ is never reached in the execution of $S_{\new}$. 
	\item It is $C_{\text{obs}}[v] = C_{\new}[v]$
	\item It is $C_{\text{obs}}[v] = C_t[v]$
	\item It is $C_{\text{obs}}[v] \neq C_t[v]$ and $C_{\text{obs}}[v] \neq C_{\new}[v]$
\end{enumerate}

Case 1 occurs if there is a branching decision based on $C_{\new} \in S_{\new}$ before $t$ is reached. Case 3 occurs if $v$ stored the value of some source code constant, which is unaffected by the changes in $S_{\new}$. Case 4 occurs if $C_t[v]$ is the result of some calculation in the program. 

Case 2 occurs if $v$ indeed stored the value that the program read from its system-level input. If case 2 occurs for some $v \in \rvar(\mse)$, $\mse$ is reversible. 

Thereby the functionality of $\mse$ depends on properties of the program under test, namely, how many variables actually store parts of the system-level input, and how many source code constants or calculations generate similar values. 

\subsection{Testing}

We instantiated the \tgb with two different unit-level analysis. \klee and a unit-level fuzzer. 

\subsubsection{\klee}

\klee\cite{Cadar2008klee} treats the parameters of the unit tests as symbolic values, and uses symbolic reasoning to come up with concrete values that explore previously unknown program path or trigger bugs. \klee cannot deal with floating-point arithmetic or symbolic pointers, it handles integers and byte sequences only. Thereby we restricted the \tgb to those data types. 

\subsubsection{Unit-level Fuzzer}

We implemented a simple unit-level fuzzer to generate new values for all parameters. We execute the unit tests with those new parameters and report all cases where the unit test fails or covers previously uncovered code. 

For integer and double types, our fuzzer uses bitflips, random values and the values \<0>, \<INT\_MAX> and \<INT\_MIN>. For strings, our fuzzer uses bitflips, sequences of random bytes, sequences of random ASCII characters, sequences of the nul byte, and sequences of \<0xFF>. Also, we implemented a mutator which takes the original string and repeats sequences thereof. 

\section{Evaluation}
\label{sec:evaluation}

In our evaluation, we attempt to answer the following research questions:
\begin{enumerate} 
	\item How do the \emph{system tests} generated by \basilisk and \fuzzilisk compare against system tests from \radamsa (\Cref{sec:eval-coverage})?
	\item Does unit-level fuzzing save time, in comparison to system-level fuzzing (\Cref{sec:eval-speedup})?
\end{enumerate}

\subsection{Subjects}

We applied our prototype implementation on six subjects:

\begin{itemize}
	\item Four of the subjects are part of \emph{GNU coreutils}, a collection of standard command line tools which is used, e.g. on Linux. The \<cut> program reads text from a file and outputs substrings, as specified by the user. The \<paste> program can be used to merge lines from different text files. The \<tac> command reads a file and outputs it in reversed order. \<b2sum> computes a message digest, some kind of checksum, from an input file. 
	
	\item \<sed> is a stream editor that applies a list of user-specified commands on its input and outputs the resulting text.
	
	\item The last subject is the \<dc> program. \<dc> is a programming languages with arbitrary-precision floating-point arithmetic. \<dc> uses reverse polish notation, where the operator follows its operands: \mbox{\<"1 2 +\char34{}>} yields the output \<3>. \<dc> has registers, which can be used as variables. 
\end{itemize}

In \Cref{tab:subjects}, we report the lines of code for each subject. Especially for the programs from \emph{GNU coreutils}, the source code repositories do contain additional code. We counted only lines of code that are in functions that are reachable from the main method of the respective program. 

\newcommand{\mcrot}[4]{\multicolumn{#1}{#2}{\rlap{\rotatebox{#3}{#4}~}}} 
\newcommand{\rot}[1]{\multicolumn{1}{l}{\rlap{\rotatebox{90}{#1}~}}} 
\begin{table}
	\caption{Evaluation subjects}
	\label{tab:subjects}
	\centering
	\sffamily
	\rowcolors{1}{Apricot}{white}
	\begin{tabularx}{\columnwidth}{l *3r X}
		\textbf{Subject} & \textbf{LoC} & \textbf{Functions} &  \\
		\hline
		b2sum & 1228 & 115 & checksum calculation \\
		paste & 662 & 79 & text processor \\
		tac & 987 & 111 & text processor \\
		bc & 3456 & 151 & arbitrary-precision calculator \\
		dc & 1997 & 136 & arbitrary-precision calculator \\
		cut & 1346 & 127 & text processor \\
		sed & 2715 & 215 & text processor \\
	\end{tabularx}
\end{table}

\subsection{Running \basilisk, \fuzzilisk and \radamsa}
\label{sec:running}

Our \basilisk and \fuzzilisk prototypes start with a set of \emph{seed tests.} First of all, they uses \radamsa to generate 10~additional system tests per seed test. Then, they carves unit tests from all system tests. Afterwards we select the unit test for the function that still contains most uncovered code, and the process of parameterizing, unit-level analysis and lifting is applied to this test. Once this is done, the next unit test will be selected and processed.  Generated system tests are executed immediately. 

In our experiments, we used a time limit of 15 minutes. Our prototype runs each system test directly after creating it, so the output is coverage information. \radamsa only generates test inputs. That means, comparing directly is not fair. While, for our tool, the time limit includes test executions, \radamsa needs additional time to execute the generated system tests. 

In order to mitigate this difference, we had \radamsa generate system tests in batches of $10$ tests each and executed each batch before generating the next one, until the time limit was exceeded. This means that for \radamsa, as for \basilisk and \fuzzilisk, system test execution time is included in the time limit.

\subsection{System Testing}
\label{sec:eval-coverage}

\begin{table*}
	\caption{Coverage reached by system tests for all tools.}
	\label{tab:coverage}
	\sffamily
	\rowcolors{1}{white}{Goldenrod}
	\centering
	\begin{tabular}{lrrrr}

\cellcolor{Goldenrod} &\cellcolor{Goldenrod}  & \cellcolor{Goldenrod} &\cellcolor{Goldenrod} & \cellcolor{Goldenrod} Improvement \\
\multirow{-2}{*}{\cellcolor{Goldenrod}Subject} 
&  \multirow{-2}{*}{\cellcolor{Goldenrod}\basilisk} 
& \multirow{-2}{*}{\cellcolor{Goldenrod}\fuzzilisk} 
& \multirow{-2}{*}{\cellcolor{Goldenrod}\radamsa} &  \cellcolor{Goldenrod} (Basilisk) \\
\midrule
b2sum   &   41.57\% &    30.60\% &  34.87\% &                   1.19 \\
paste   &   41.93\% &    38.32\% &  37.28\% &                   1.12 \\
tac     &   33.55\% &    31.98\% &  29.92\% &                   1.12 \\
dc      &   19.13\% &    19.13\% &  35.73\% &                   0.54 \\
cut     &   22.58\% &    21.36\% &  21.61\% &                   1.05 \\
sed     &   30.62\% &    24.36\% &  19.24\% &                   1.59 \\

\end{tabular}

\end{table*}

In this section, we answer research question~1, namely, how the \tgb compares to another system test generator. 

We ran \fuzzilisk, \basilisk and \radamsa as described in \Cref{sec:running}, with a timeout of 15~minutes. We repeated each run 5~times, with 5~different random seeds. For \fuzzilisk, both the unit-level analysis and system-level fuzzing is randomized. For \basilisk, unit-level analysis is deterministic, but system-level testing is still randomized.  For all subjects, we measure the branch coverage over time.

\subsubsection{Coverage}

\Cref{tab:coverage} summarizes the branch coverage obtained, as a mean over five runs.  In all subjects but \<dc>, the coverage achieved by \basilisk improves over \radamsa alone, showing a clear benefit of the bridging of system-level and unit-level test generators.

\begin{result}
	For all subjects but \<dc>, the \basilisk bridge of system-level test generation (\radamsa) and unit-level test generation (\klee) increases coverage over \radamsa alone between~5\%~(\<cut>)~and~59\%~(\<sed>).
\end{result}

\<dc>, being an arbitrary-precision calculator, stores numbers as strings of bytes, which our implementation cannot associate with input fragments and thus neither map nor lift.  This claim is further supported by \Cref{tab:functions}, showing that \basilisk analyzes just one unit test for \<dc>. All other tests had no parameterizable inputs. 

\begin{table}
	\caption{Functions covered in unit tests.}
	\label{tab:functions}
	\sffamily
	\rowcolors{1}{Goldenrod}{white}
	\centering
	\begin{tabular}{lrrr}

Subject & \# functions &  \basilisk &  \fuzzilisk \\
\midrule
b2sum   &          204 &        17 &         12 \\
paste   &          152 &        21 &          9 \\
tac     &          177 &        18 &          9 \\
dc      &          154 &         1 &          3 \\
cut     &          205 &        37 &          7 \\
sed     &          290 &        38 &         12 \\

\end{tabular}

\end{table}

\begin{result}
	Non-standard data representations of strings and numbers, as in \<dc>, need special implementations for mapping and lifting.
\end{result}

\subsubsection{Coverage Over Time}

\begin{figure}
	\caption{Coverage over time for \basilisk and \radamsa}
	\label{fig:cov-vs-time}
	\centering
	\includegraphics[width=.9\columnwidth]{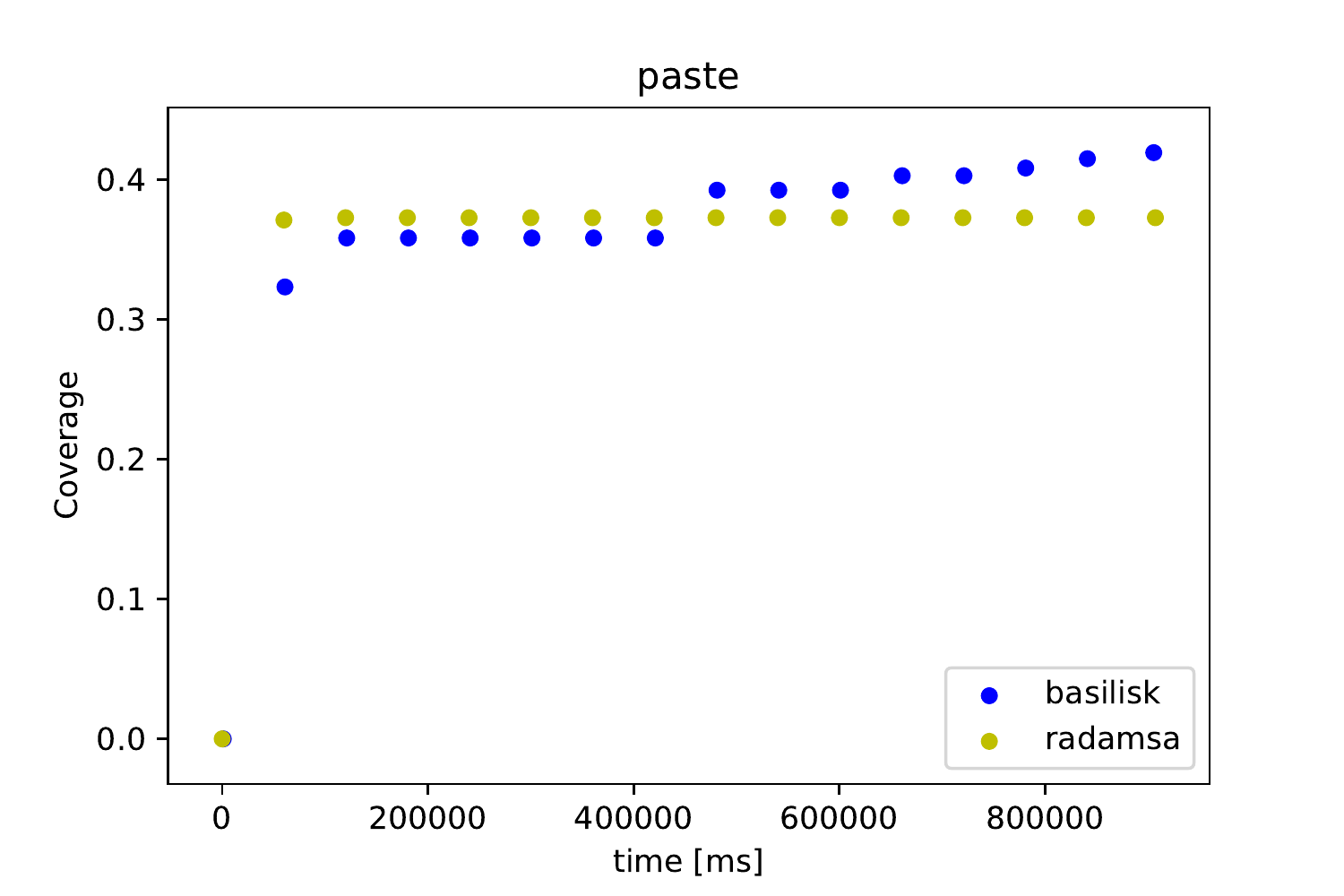}
\end{figure}

Does the increased coverage also translate into time savings?  \Cref{fig:cov-vs-time} shows a plot of the branch coverage achieved over time for \<paste>.  The diagrams list time on the x-axis, and the coverage achieved so far by \basilisk or \radamsa respectively on the y-axis.  This is a typical representative; the other subjects show similar plots. 

\basilisk initially has worse coverage than \radamsa. This is due to the extra overhead of carving.  However, for all subjects, there is a point in time when \basilisk outperforms \radamsa.  Typically, this happens at about 8~minutes.   Note that \basilisk's coverage after 8~minutes also outperforms \radamsa coverage after 15~minutes.  This means that with \basilisk, one can achieve the same coverage in about half of the time.  Again, \<dc> is the exception.

\begin{result}
	In our experiments, \basilisk achieves the same coverage as \radamsa in about half the time.
\end{result}

While the increase typically is slow and steady, there is a massive jump in \<sed>. \<sed> accepts an input language, which consists of several commands. The jump happens when \basilisk (or to be precise, \klee) discovers the \<q> command, which terminates \<sed>. It proceeds to generate a lot of tests which contain \<q> and thereby covers a lot of new code. \radamsa did not manage to trigger the \<q> command in our experiments. 

\subsubsection{Symbolic Testing vs. Random Testing}

The \fuzzilisk combination of \radamsa and a unit-level random fuzzer outperforms \radamsa only on \<paste> and \<tac>, and is never better than the \basilisk combination of \radamsa and \klee; it also covers fewer functions.  This shows a clear benefit of the \tgb, namely that it allows to apply symbolic testing techniques at the unit level.

\begin{result}
	At the unit level within the \tgb, symbolic~testing (\basilisk with \klee) consistently covers~more~code than random~testing (\fuzzilisk).
\end{result}

\subsubsection{Lifting Performance}

\begin{table*}
	\caption{Unit Tests selected for lifting in \fuzzilisk}
	\label{tab:lifting_fuzzilisk}
	\sffamily
	\rowcolors{1}{Apricot}{white}
	\centering
	\begin{tabular}{lrrrr}

Subject & \# Unit Tests & \# Lifted Tests & \% lifted & \% effective \\
\midrule
b2sum   &   9005075.20 &          53.40 &  0.0007\% &     0.0001\% \\
paste   &   4654504.60 &          37.60 &  0.0008\% &     0.0002\% \\
tac     &   2543487.80 &          35.20 &  0.0014\% &     0.0003\% \\
dc      &    475426.60 &          14.20 &  0.0030\% &     0.0000\% \\
cut     &   1167711.00 &          29.60 &  0.0026\% &     0.0002\% \\
sed     &   4656782.00 &         105.20 &  0.0023\% &     0.0008\% \\
total   &   3750497.87 &          45.87 &  0.0012\% &     0.0003\% \\

\end{tabular}

\end{table*}

\begin{table}
	\caption{Effectiveness of lifting in \basilisk}
	\label{tab:lifting_basilisk}
	\sffamily
	\rowcolors{1}{Goldenrod}{white}
	\centering
	\begin{tabular}{lrr}

Subject & \# Lifted Tests & \% effective \\
\midrule
b2sum   &         876.40 &     3.9049\% \\
paste   &         757.80 &     3.6158\% \\
tac     &         870.60 &     2.2169\% \\
dc      &         266.60 &     0.0000\% \\
cut     &         393.40 &     2.5420\% \\
sed     &         670.00 &    12.3893\% \\
totals  &         639.13 &     4.5322\% \\

\end{tabular}

\end{table}

\Cref{tab:lifting_fuzzilisk} shows how many tests were lifted to the system level within \fuzzilisk. This number is rather low, less than 1\% of the tests in total. The percentage of effective lifts is even lower. We counted a test as \emph{effective} if it increases coverage in the program under test. This means that \fuzzilisk in fact tries many execution paths that do not lead to new coverage, or crashes. It would be impractical to test all those execution paths with system tests.  These numbers shed a light on the usefulness (or better: non-usefulness) of unguided random tests at the function level, and motivate their lifting into a system test to validate their results.

\begin{result}
	In our experiments with \radamsa and a unit-level fuzzer, less than 0.001\% of random function invocations that increase coverage or cause failures also do so when lifted back to the system level.
\end{result}

Being a symbolic tester, \klee always executes multiple program path at the same time, and only reports those that trigger a bug or cover new code. Thereby, the full number of unit tests, as given in \Cref{tab:lifting_fuzzilisk}, is not available for the combination of \radamsa and \klee (which we call \basilisk). \Cref{tab:lifting_basilisk} thereby only gives the number of lifted tests (path reported by \klee), and the effectiveness. 

Also, \basilisk gives a different picture here: Except for \<dc>, all subjects show a lifting effectiveness above 2\%, with a maximum of 12\% for \<sed>. This again provides evidence for the hypothesis that \fuzzilisk's performance is mainly a problem with the unit-level fuzzing tool. 

\begin{result}
	In our experiments with \radamsa and \klee, 4.53\% of paths at unit level that increase coverage or cause failures also do so when lifted back to the system level.
\end{result}


\subsection{Unit Testing}
\label{sec:eval-speedup}

\begin{table}
	\caption{Speed advantage of System Tests vs. Unit Tests}
	\label{tab:speedup}
	\sffamily
	\rowcolors{1}{GreenYellow}{white}
	\centering
	\begin{tabular}{lrrr}

Subject & \multicolumn{2}{c}{Median Runtime[ms]} & Mean \\
 & System Tests & Unit Tests & speed-up \\
\midrule
b2sum   &                              39 &                             2 &       1165.00 \\
paste   &                             114 &                             0 &       1196.30 \\
tac     &                             160 &                             4 &       1393.43 \\
dc      &                            1620 &                             0 &       5005.40 \\
cut     &                              32 &                             0 &        281.21 \\
sed     &                              98 &                             0 &       1684.92 \\
total   &                              73 &                             0 &        495.06 \\

\end{tabular}

\end{table}

Unit tests typically run much faster than system tests, as they execute much less code. \Cref{tab:speedup} shows how much faster they are within our \fuzzilisk prototype. On average, a unit test is \textbf{495$\times$~faster} than a system test for the same program.

\begin{result}
	A \tgb allows to focus test generation on selected unit tests, which run several orders of magnitude faster than the original system test.
\end{result}

This means that if we want to test one single function only (say, because it is recently changed), we can save a huge amount of time, or spend the same time on more thorough testing.  

For the example in \Cref{sec:example}, a single attempt to login on the system level takes about 1.7 seconds, while a single unit test takes only 37 milliseconds. In comparison to other subjects, 37 milliseconds are still a lot, but then, this subject executes an SQL query as part of the unit test. 
For symbolic testing, keep in mind that the values in this example travel via a HTTP request, through a python layer into the database, written in C. We are unaware of any symbolic tool which can handle the entire stack. 

\section{Limitations and Threats to Validity}

Threats to external validity concern our ability to generalize the results of our study. We cannot claim that the results of our experimental evaluation are generalizable. A concern is that we establish mappings from inputs to unit-level values via (sub-)string equality. In our django example, the \tgb can discover the user name, but not the password. The password does appear as a parameter to a \<sqlite3\_bind\_text()> call, but it has been processed (hashed and salted) before.  This means we can not recognize the value, and will not mark it as a parameter.  Our technique thus primarily works on functions with a one-to-one mapping to input.  However, in practice we often see that inputs and options go unchanged through several layers of a program, before they are finally processed, as it happens with the user name in our example.

The quality of carved unit tests depends on the choice of seed system tests. A good seed test gives the \tgb more chances to carve out unit tests, and it also gives \radamsa more opportunities to mutate system-level inputs, again yielding better carved unit tests.  We do not see system test generation and unit test generation as alternatives, but as complementary, which is the point of this work.

\section{Related Work}
\label{sec:related-work}

\paragraph{Carving Unit Tests}

Carving unit tests was introduced by Elbaum et al.~\cite{elbaum2009carving} as a means to speed up repeated system tests, for instance in the context of regression testing.  His work used the Java infrastructure, whereas our mechanism carves \emph{parameterized unit tests}~\cite{tillmann2005parameterized}, where function parameters whose values can be mapped to system input are left as parametric and thus open for unit test generators to explore.  To the best of our knowledge, \basilisk~is the first tool to implement parametric unit test carving; it also is the first tool to carve unit tests from \<C> programs.

\paragraph{Extracting C Memory Snapshots}

Interpreting and recovering \<C> data structures at runtime is notoriously difficult, since every programmer can implement not only their own data structures, but also her individual memory management. The work of Zimmermann and Zeller~\cite{Zimmermann2001graphs} on extracting and visualizing \<C> runtime data structures (``memory graphs'') as well as their later application in  debugging~\cite{Zeller2002causeEffect, Cleve2005causes, Polishchuk2007inference} is related to ours in that all these works attempt to obtain a reliable and reproducible snapshot of \<C> data structures.  Yet, all these works use these data structures for the purposes of debugging and program understanding rather than carving.

\paragraph{Generating System Tests}

The idea of \emph{generating software tests} is an old one: To test a program~$S$, a producer~$P$ will generate inputs for~$S$ with the intent to cause it to fail.  To find bugs, a producer need not be very sophisticated; as shown in the famous ``fuzzing'' paper of~1989, simple random strings can quickly crash programs~\cite{miller1990}.  

To get deeper than scanning and parsing routines, though, one requires syntactically correct inputs.  To this end, one can use formal specifications of the input language to generate inputs---for instance, leveraging \emph{context-free grammars} as producers~\cite{purdom1972}.  The \textsc{Langfuzz} test generator~\cite{Holler2012langfuzz} uses its grammar for \emph{parsing} existing inputs as well and can thus combine existing with newly generated input fragments.

Today's most popular test generators take \emph{input samples} which they mutate in various ways to generate further inputs.   \emph{American Fuzzy Lop,} or AFLFuzz, combines mutation with search-based testing and thus systematically maximizes code coverage~\cite{zalewski2018aflfuzz}.  More sophisticated fuzzers rely on \emph{symbolic analysis} to automatically determine inputs that maximize coverage of control or data paths~\cite{Godefroid2008whitebox}. The \klee~tool~\cite{Cadar2008klee} is a popular symbolic tester for \<C>~programs. 

In our experiments, we use \radamsa~\cite{helin2018radamsa} which applies a number of mutation patterns to systematically widen the exploration space from a single input.  Instead of \radamsa, any other system-level test generator could also do the job.

\paragraph{Generating Unit Tests}

The second important class of testing techniques works at the \emph{unit level,} synthesizing calls of individual functions.  These techniques separate in two branches: \emph{random} and \emph{symbolic}.  

\emph{Random} tools operate by generating random function calls, which are then executed.  A typical representative of this class is the popular \textsc{Randoop}~\cite{pacheco2007randoop} tool.  Random calls can be systematically refined towards a given goal: EvoSuite~\cite{Fraser2011evolution} uses a search-based approach to evolve generated call sequences towards maximizing code coverage.

\emph{Symbolic} techniques symbolically solve \emph{path conditions} to generate inputs that reach as much code as possible. PEX~\cite{tillmann2008pex} fulfills a similar role for .NET programs, working on \emph{parametrized unit tests} in which individual function parameters are treated symbolically.

Compared to the system level, test generation at the unit level is very efficient, as a function call takes less time than a system invocation or interaction; furthermore, exhaustive and symbolic techniques are easier to deploy due to the smaller scale.  The downside is that generated function calls may lack realistic context, which makes exploration harder; and function failures may be false alarms because of violated implicit preconditions.
Kim et al.\cite{kim2018precise} devised a technique to use the static calling context of a function in unit-testing. They report high bug detection ability and high precision, but still some false positives. In contrast, we derive the context from execution, that is, with a dynamic analysis. Also, validating all unit-level failures at the system level means that we can recover from false alarms and any remaining failures are true failures.

Godefroid\cite{Godefroid2014microexec} executes individual units without context, and generates context on the fly if it is accessed. However, they use no information on the actual context, as we do. Thereby, they state "Micro execution of a code fragment makes sense mostly if all its inputs are unconstrained" (Section 4).

\paragraph{Generating Parameterized Unit Tests}

A number of related works has focused on obtaining parameterized unit tests by starting from existing or generated unit tests.  Retrofitting of unit tests~\cite{Thummalapenta2011retrofitting} is an
approach where existing unit tests are converted to parameterized unit
tests, by identifying inputs and converting them to parameters.  The technique of Fraser and Zeller~\cite{Fraser2011gpu} starts from concrete inputs
and results, using test generation and mutation to systematically
generalize the pre- and postconditions of existing unit tests.  The recently presented \emph{AutoPUT} tool~\cite{Tsukamoto2018autoPUT} generalizes over a set of related unit tests to extract common procedures and unique parameters to obtain parametrized unit tests.  In contrast to all these works, our technique carves parameterized unit tests directly out of a given run, identifying those values as parameters that are present in system input.

\section{Conclusion}
\label{sec:conclusion}

A \tgb turns a system test into a set of parameterized unit tests, which can all individually be tested by a unit-level test generator.  The bridge allows arbitrary wide gaps between system input and unit invocations, arbitrary combinations of system test generators and unit test generators, and avoids false positives at the unit level via mapping and lifting.  In our experiments, the combination of the \radamsa  system-level test generator and the \klee unit-level test generator improved coverage over \radamsa in all programs with standard integer and string representations.  Focusing testing on recently changed or otherwise error-prone functions brings the potential for important time savings during testing. 

We showed, in an example, how the \tgb helps to span the distance between HTTP requests and the database backend of a web application. 
In our future work, we will be exploring additional applications and improvements for the \tgb.  These include:

\textbf{Action-based carving.} In this paper, we used state-based carving. That means that the execution context for a carved unit test is a replication of the memory observed in a program run. Action-based carving instead records the interactions that created this memory state, and uses the same function calls to recreate the memory for the test. This may simplify carving, as it is no longer necessary to dump out memory state, functions and the memory locations they operate on suffice. However, it may lead to longer, more complicated unit tests, which would be harder to analyze at the unit level. We may further investigate this relationship.

\textbf{Dynamic taints.}  Besides simple value-based mapping of strings and numbers, we are exploring leveraging \emph{dynamic taints} to track input characters throughout program execution.  Lifting failure-inducing values back to the system level may then demand inverting processing functions, which can be a challenge.

\textbf{Grammar-based testing.}  To further improve mapping and lifting, we are working on using grammar-based test generators at the system level, allowing us to precisely associate a language element with a failure-inducing function argument.  Producing from a grammar allows us to reuse these values in many more contexts, creating synergies between system-level and unit-level test generation.

\bibliography{main.bib}
\bibliographystyle{plain}
\end{document}